\documentclass[%
 reprint,
 superscriptaddress,
 showkeys,
 amsmath,amssymb,
 aps,
 floatfix,
]{revtex4-1}

\usepackage{graphicx}% Include figure files
\usepackage{dcolumn}% Align table columns on decimal point
\usepackage{bm}% bold math
\usepackage{hyperref}% add hypertext capabilities

\newcommand{\Tg}{T_g}
\newcommand{\Tdep}{T_\text{dep}}
\newcommand{\sbb}{\sigma_{bb}}
\newcommand{\stt}{\sigma_{tt}}
\newcommand{\sbt}{\sigma_{bt}}
\newcommand{\ebb}{\epsilon_{bb}}
\newcommand{\ett}{\epsilon_{tt}}
\newcommand{\ebt}{\epsilon_{bt}}

\begin{document}

\title{Effects of Microstructure Formation on the Stability of Vapor Deposited Glasses}

\author{Alex R. Moore}
\affiliation{Department of Chemical and Biomolecular Engineering, University of Pennsylvania, Philadelphia, Pennsylvania 19104, USA}
\author{Patrick J. Walsh}%
\affiliation{Department of Chemistry, University of Pennsylvania, Philadelphia, Pennsylvania 19104, USA}
\author{Zahra Fakhraai}%
\affiliation{Department of Chemical and Biomolecular Engineering, University of Pennsylvania, Philadelphia, Pennsylvania 19104, USA}
\affiliation{Department of Chemistry, University of Pennsylvania, Philadelphia, Pennsylvania 19104, USA}
\author{Robert A. Riggleman}%
\email{rrig@seas.upenn.edu}
\affiliation{Department of Chemical and Biomolecular Engineering, University of Pennsylvania, Philadelphia, Pennsylvania 19104, USA}

\date{\today}

\begin{abstract}
Glasses formed by physical vapor deposition (PVD) are an interesting new class of materials, exhibiting properties thought to be equivalent to those of glasses aged for thousands of years. Exerting control over the structure and properties of PVD glasses formed with different types of glass-forming molecules is now an emerging challenge. In this work, we study coarse grained models of organic glass formers containing fluorocarbon tails of increasing length, corresponding to an increased tendency to form microstructures. We use simulated PVD to examine how the presence of the microphase separated domains in the supercooled liquid influences the ability to form stable glasses. This model suggests that increasing molecule tail length results in decreased thermodynamic and kinetic stability of the molecules in PVD films. The reduced stability is further linked to the reduced ability of these molecules to equilibrate at the free surface during PVD. We find that as the tail length is increased, the relaxation time near the surface of the supercooled equilibrium liquid films of these molecules are slowed and become essentially bulk-like, due to the segregation of the fluorocarbon tails to the free surface. Surface diffusion is also markedly reduced due to clustering of the molecules at the surface. Based on these results, we propose a trapping mechanism where tails are unable to move between local phase separated domains on the relevant deposition time scales.
\end{abstract}

\maketitle

\section{\label{sec:intro}Introduction}

Amorphous films of small organic molecules are widely used in applications ranging from organic electronics \cite{shirota2007charge, tang1987organic, salbeck1997low} to protective coatings \cite{mahltig2005optimized} to nano-imprint lithography \cite{pires2010nanoscale}. Recent studies have shown that glass packings of organic molecules created using physical vapor deposition (PVD), while the deposition substrate temperature ($\Tdep$) is held below the material's glass transition temperature ($\Tg$), demonstrate remarkable kinetic and thermodynamic stability \cite{swallen2007organic, kearns2008hiking, leon2010stability, fakhraai2011structural, ishii2008anomalously, dalal2012density, yu2013ultrastable, sepulveda2013manipulating}. These stable glasses are thought to utilize the enhanced mobility at the free surface of the film, such that molecules are able to sample additional configurations and find optimal local positioning before being constrained by subsequent layers \cite{zhu2011surface, brian2013surface, chua2015much, zhang2016long, zhang2017invariant}. This surface-mediated equilibration (SME) results in high density, low energy films thought to be equivalent to those that have been aged for hundreds or thousands of years. Substrate temperature (relative to $\Tg$) and deposition rate are integral to the surface dynamics of this process and therefore can have a large effect on the properties of PVD glass films \cite{dalal2013high, kearns2007influence}. 

%%%

In light of the discovery of these stable glasses, many have turned to molecular dynamics (MD) simulations to further study these systems in microscopic detail. Singh and de Pablo have developed an algorithm to computationally mirror the vapor deposition process that is able to create the expected lower energy, higher stability films \cite{singh2011molecular}. This protocol has since been used in simple model \cite{singh2013ultrastable, lyubimov2013model, lin2014molecular} and atomistic studies \cite{singh2011molecular, antony2017influence} alike. Lyubimov {\it et al}. also used MD simulations to demonstrate the link between a molecule's equilibrium liquid surface properties and the orientation of the molecules during the deposition process, by observing trends in anisotropy of the molecules near the surface \cite{lyubimov2015orientational} that were in good agreement with experiments \cite{dalal2015tunable}.

%%%

Predicting and controlling the properties of PVD glasses formed with different types of small organic molecules remains a fundamental challenge. In particular, intermolecular interactions seem to have a large effect on the stability of PVD films \cite{souda2010structural, sepulveda2012glass, kasina2015dielectric, tylinski2016vapor}. This was demonstrated recently by Laventure {\it et. al.}, who showed that model triazine derivatives with increasing hydrogen bonding capability exhibit lowering kinetic stability \cite{laventure2017influence}. This, along with other recent work \cite{chen2016hydrogen, zhang2015fast, zhang2016surface, chen2017influence}, suggest that hydrogen bonding could inhibit the formation of stable glasses, potentially due to limited surface mobility. However, much remains to be learned about the exact nature of the effects of intermolecular forces in general, and the resulting microstructures they can create upon PVD, or even to what extent these microstructures can be imprinted without losing stability. Further studying these effects could elucidate key mechanistic details behind the SME process, allow for the substantial improvement of PVD-engineered glass film properties, and broaden the classes of molecules that can be employed to generate stable PVD glasses.

%%%

\begin{figure}[ht]
\centering
\includegraphics[width=0.7\linewidth]{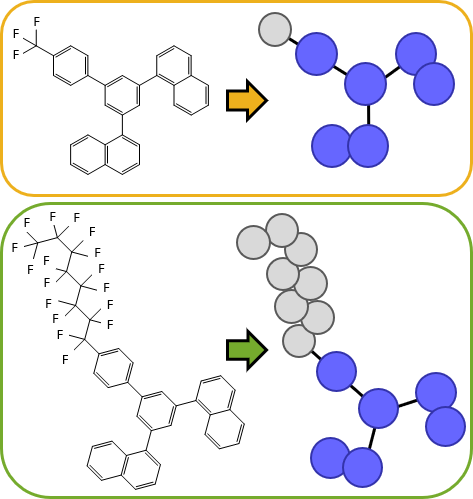}
\caption{Molecular representation of the 1-tail and 8-tail organic glass formers (left) and corresponding coarse-grained models (right). Body Lennard-Jones particles are colored blue and tail particles are colored gray. }
\label{fig:coarse}
\end{figure}

In this work, we aim to systematically examine the effects of intermolecular interactions and the propensity for microstructure formation on the surface mobility and stability of simulated PVD glasses. To gain insight on the local structure and mechanism, we use MD simulations to study empirical coarse grained models of organic molecules containing model fluorocarbon tails of increasing length: 0, 1, 4, and 8 fluorocarbons. The organic glass-formers of interest here, shown in Figure \ref{fig:coarse}, are made up of two distinct sections: the phenyl "body" and the fluorocarbon "tail". These groups exhibit vastly different dispersion forces, and thus locally each component will avoid mixing with each other. Our coarse-grained model represents each benzene and each fluorocarbon as one of two distinct Lennard-Jones sites, also shown in Figure \ref{fig:coarse}. By varying the length of the "tail" section, we can tune the degree to which these molecules are able to micro-separate, and then examine the impact this has on the resulting PVD film properties. We note that although the energy and length scale of the interactions were chosen based on the relative polarizabilities and sizes of the phenyl and fluorinated carbon groups, no effort was made to quantitatively map between a higher resolution (e.g., atomistic) model. 

In the results presented below, we observe a clear trend indicating decreasing PVD film stability and increasing microstructure formation with increasing tail length. We then study the dynamic properties of these model glass-formers, namely the mean squared displacement (MSD) at the free surface and the relaxation time of the equilibrium supercooled liquid as a function of depth from the free surface. These results lead us to propose a trapping mechanism in which the longer tail molecules are less likely to escape from locally phase-separated domains at the surface, which is supported by the short-time behavior of these molecules just after deposition and observed trends in stability at slower deposition rates. Furthermore, the longer tails tend to segregate to the free surface, thus slowing the dynamics at the layers right below the free surface where the tails are depleted. The tail segregation effect is also observed in the PVD films, but surprisingly do not lead to bi-layer formation as molecular orientation and rearrangement further below the free surface reduces the degree of segregation and prevents the formation of layers.

\section{\label{sec:sim}Simulation Details}

For the coarse grained models used in this study, each phenyl or $\text{CF}_2$/$\text{CF}_3$ group in the molecule of interest was represented by a sphere interacting with a Lennard-Jones potential. To represent the different sizes and polarizabilities of the ``body'', $b$, groups and the ``tail'', $t$, groups, Lennard-Jones parameters were chosen $\sbb = 1.0$, $\ebb = 1.0$, $\stt = 0.6$, and $\ett = 0.05$, which are consistent with the relative sizes and polarizabilities of the constituents. For the cross interactions between $b$ and $t$ components, standard Lorentz-Berthelot mixing rules were applied such that $\sbt = (\sbb + \stt) / 2$ and $\ebt = \sqrt{\ebb \ett}$. The substrate was generated by taking a slice of a disordered Lennard-Jones system of density $\approx$ 1, and substrate interactions were neutral with both $b$ and $t$ type particles. For each molecule, $T_g$ was determined by measuring the potential energy during constant pressure cooling ramps of bulk systems of 1000 molecules in a box with periodic boundaries in the $x$, $y$, and $z$ directions, and identifying the temperature at which the thermal expansion coefficient changed. For 0-tail and 1-tail molecules, $T_g = 0.55$, for the 4-tail molecule, $T_g = 0.48$, and for the 8-tail molecule, $T_g = 0.41$. The PVD process was simulated using a method similar to the one developed by Lyubimov {\it et. al.} \cite{lyubimov2013model} The time allowed for a deposited molecule to cool on the surface of the film was varied to achieve the effect of different deposition rates. A ``normal'' deposition rate allowed 150 $\tau$ and a ``slow'' deposition rate allowed 300 $\tau$. To create the equilibrium liquid, films created via the PVD protocol were heated to well above their glass transition temperature, $T = 1.5T_g$, allowed to relax, and subsequently cooled to $T = 1.2T_g$. More details of the simulation methods can be found in the Materials and Methods section. 

\begin{figure}[ht]
\centering
\includegraphics[width=0.7\linewidth]{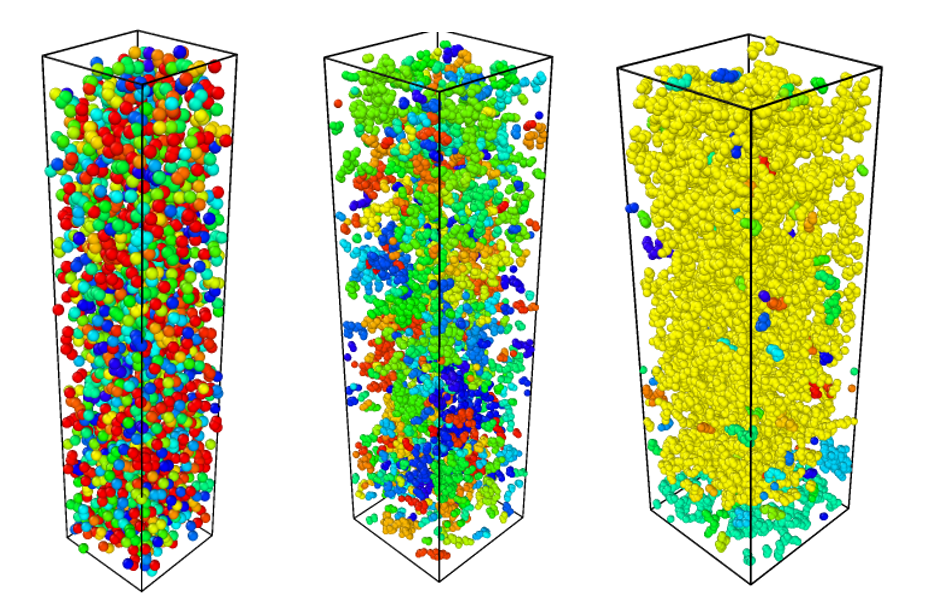}
\caption{Visualizations of the tail particles in the aggregate analysis performed on PVD films of 1, 4, and 8-tail systems (left to right). Each of these films was deposited at 0.75$T_g$ at the typical 150 $\tau$ deposition rate. Aggregates are defined as tail particles within 1 $\sbb$ of each other. Color is generated by a randomly assigned aggregate number.}
\label{fig:aggregate}
\end{figure} 

\section{\label{sec:results}Results} 

\subsection{\label{sec:struct}Structure and stability of PVD Glasses} 

Several films of the model fluorocarbon tail molecules (0,1,4, and 8-tail) were deposited using the algorithm outlined in the Materials and Methods section and a $\Tdep$ ranging from 0.75-0.95$\Tg$. Figure \ref{fig:aggregate} demonstrates the difference in the cluster formation between molecules of varying tail lengths. Analysis of these microstructures were performed by cooling the as-deposited and transformed supercooled liquid films to 0.75$T_g$, and sorting each tail particle into a cluster, defined such that a particle was considered to be in the same cluster as another if the pair were within 1$\sbb$ of each other. Quantitatively speaking, the average cluster size across all observed films of 1-tail molecules was 1.5 tails and the largest cluster in any film represented just 0.6$\%$ of total tails in the system. For the 4-tail molecules, the average cluster contained 3.8 tails and the largest cluster represented 6.5$\%$ of the system; for the 8-tail molecules, the average cluster contained 19.1 tails and the largest cluster represented 87$\%$ of the system. No significant difference in the cluster size was observed between different deposition temperatures or rates, or between PVD and liquid-quenched glasses. Figure \ref{fig:aggregate} further shows that smaller tail molecules lack the ability to form widespread clusters, while larger molecules can arrange themselves to form near-percolating microstructures.

\begin{figure}[ht]
\centering
\includegraphics[width=0.7\linewidth]{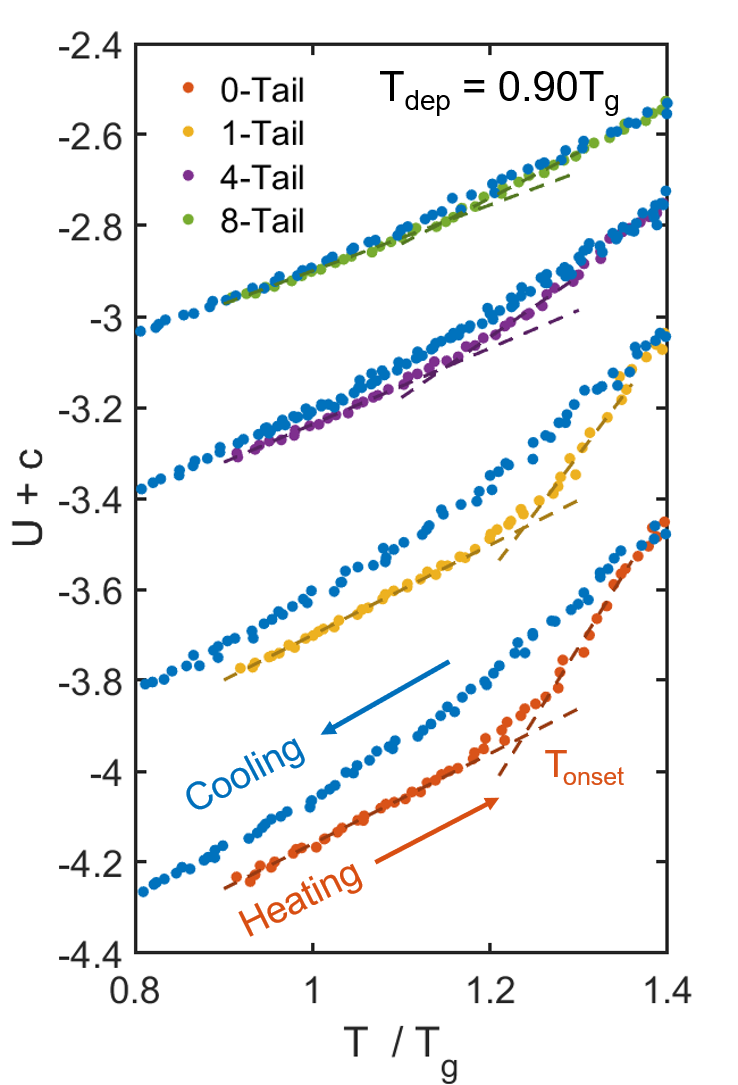}
\caption{Temperature ramp results, plotted as potential energy vs. $T / \Tg$, for PVD films of each molecule type at $\Tdep = 0.90 \Tg$ and $150 \tau$ deposition rate. A constant has been added to shift potential energy of the respective curves for ease of comparison. }
\label{fig:stable}
\end{figure}

To investigate the stability of the simulated PVD films, we used temperature ramping simulations to observe the change in the system's potential energy upon heating and subsequent cooling of as-deposited glasses with various tail lengths, as can be seen in Figure \ref{fig:stable}. We find that across the entire range of relative temperatures studied, a clear trend emerges in the film stability. As Figure \ref{fig:stable} shows, those molecules with longer tails, and thus larger clusters, exhibit smaller changes in their potential energy upon transformation to liquid-quenched glasses and demonstrably lower onset temperatures relative to $T_g$ than their shorter tail counterparts. This indicates that the differences in the interaction potentials and the resulting microstructures lead to less stable PVD glasses.

%%%

\begin{figure}[ht]
\centering
\includegraphics[width=0.75\linewidth]{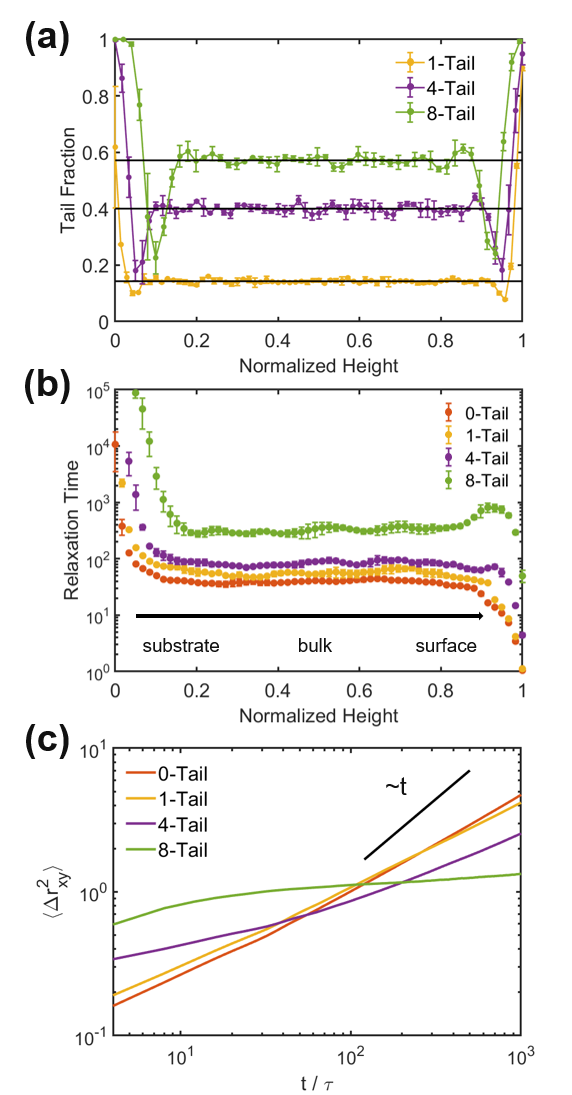}
\caption{(a) Local tail particle fraction vs. normalized height from the substrate, (b) the local relaxation time vs. normalized height from the substrate, and (c) mean squared displacement in the $x-y$ plane of the free surface layer, with solid black lines representing system average, for films of equilibrium liquid ($T = 1.2\Tg$) of each molecule. Each film is 50-60 $\sbb$ in height.
}
\label{fig:eqliq}
\end{figure} 

%%%

\subsection{\label{sec:supliq}Supercooled liquid surface properties}
To learn more about the mechanism behind the instability of PVD glasses formed with the longer tail molecules, we studied films of each molecule in the equilibrium supercooled liquid state, at the temperature $T = 1.2 \Tg$. We divided each simulated film in this equilibrium state into layers of thickness 1 $\sbb$ in the $z$-direction and measured the number of tail and body particles in each layer. Figure \ref{fig:eqliq}(a) shows the fraction of tail particles in each layer for equilibrium supercooled liquid films of the 1, 4, and 8-tail molecules. It can be seen that the tails in these films show a strong tendency to migrate to the free surface, resulting in a tail depletion zone that grows stronger in deviation from system average fraction and in length scale with increasing tail size. Thus the degree of enhanced dynamics at the free surface and its propagation depth into the film also diminishes with increased tail length as seen by the plot of relaxation times of the body particles with increasing $z$-position in the equilibrium liquid state (Figure \ref{fig:eqliq}(b)). The 8-tail molecule in particular shows bulk-like relaxation behavior up until the very top layer, in contrast to the 0-tail molecule which shows a decrease in the relaxation time that extends farther from the surface. In addition, there are local peaks in the relaxation time for the 4-tail and 8-tail molecules, and by comparing with Figure \ref{fig:eqliq}(a) we can attribute these variations in the relaxation time to the variations in the molecular packing near the interface. Given the favorable surface arrangement of the tail particles, the body particles show preference to be on the secondary layer, and this change in packing leads to longer relaxation times in this region. Figure \ref{fig:eqliq}(c) also compares the MSD along the $x-y$ plane for all particles within the free surface layer. From these results we can see that the molecules with longer tails exhibit sub-diffusive behavior, and the MSD for the 8-tail molecule exhibits a caging plateau with little discernible surface diffusion. 

\begin{figure}[ht]
\centering
\includegraphics[width=0.75\linewidth]{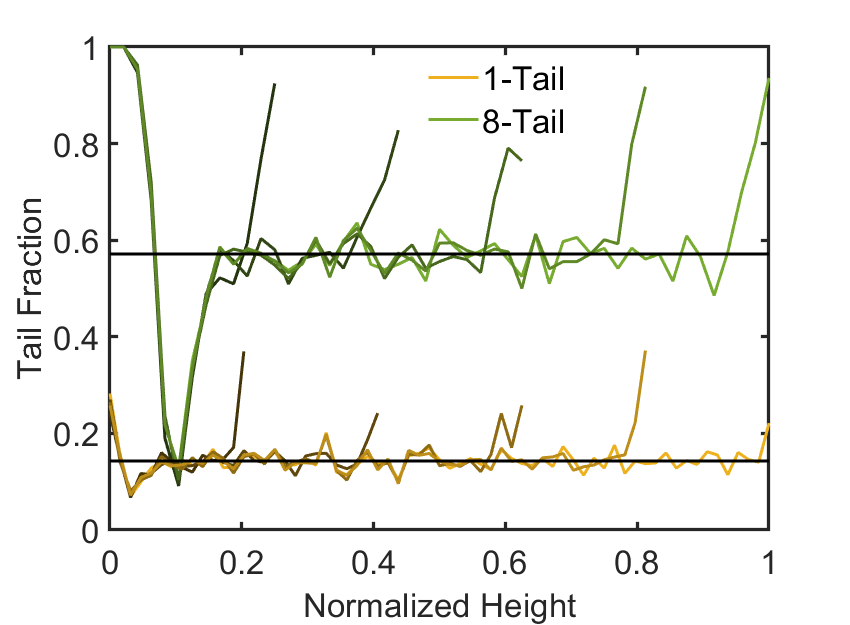}
\caption{Local tail particle fraction vs height from the substrate, with solid black lines representing system average, at $20\%, 40\%, 60\%, 80\%,$ and $100\%$ of one instance of the deposition process for the 1-tail and 8-tail molecules at $0.90\Tg$. The color of the line changes from dark to light the further along in the deposition process.}
\label{fig:tail-frac-dep}
\end{figure}

%%%

\subsection{\label{sec:pvddyn}Dynamics of the PVD glass surfaces}
The tendency for the tails to migrate to the free surface also impacts the formation of the vapor deposited glasses as can be seen in Figure \ref{fig:tail-frac-dep}. This figure shows snapshots of the film composition throughout one instance of the deposition process at 0.90$\Tg$. The tail composition is enriched at the surface layer, but as the deposition process continues, this initial layering is reconfigured.  This suggests that the liquid-like mobile layer at the film free surface extends beyond the length of these tails, even for the 8-tail molecules where the free surface dynamics are slowed. Figure S1  of the online supplementary information (SI) looks further into the displacement distribution of tails at different locations relative to the free surface.

%%%

\begin{figure}[ht]
\centering
\includegraphics[width=\linewidth]{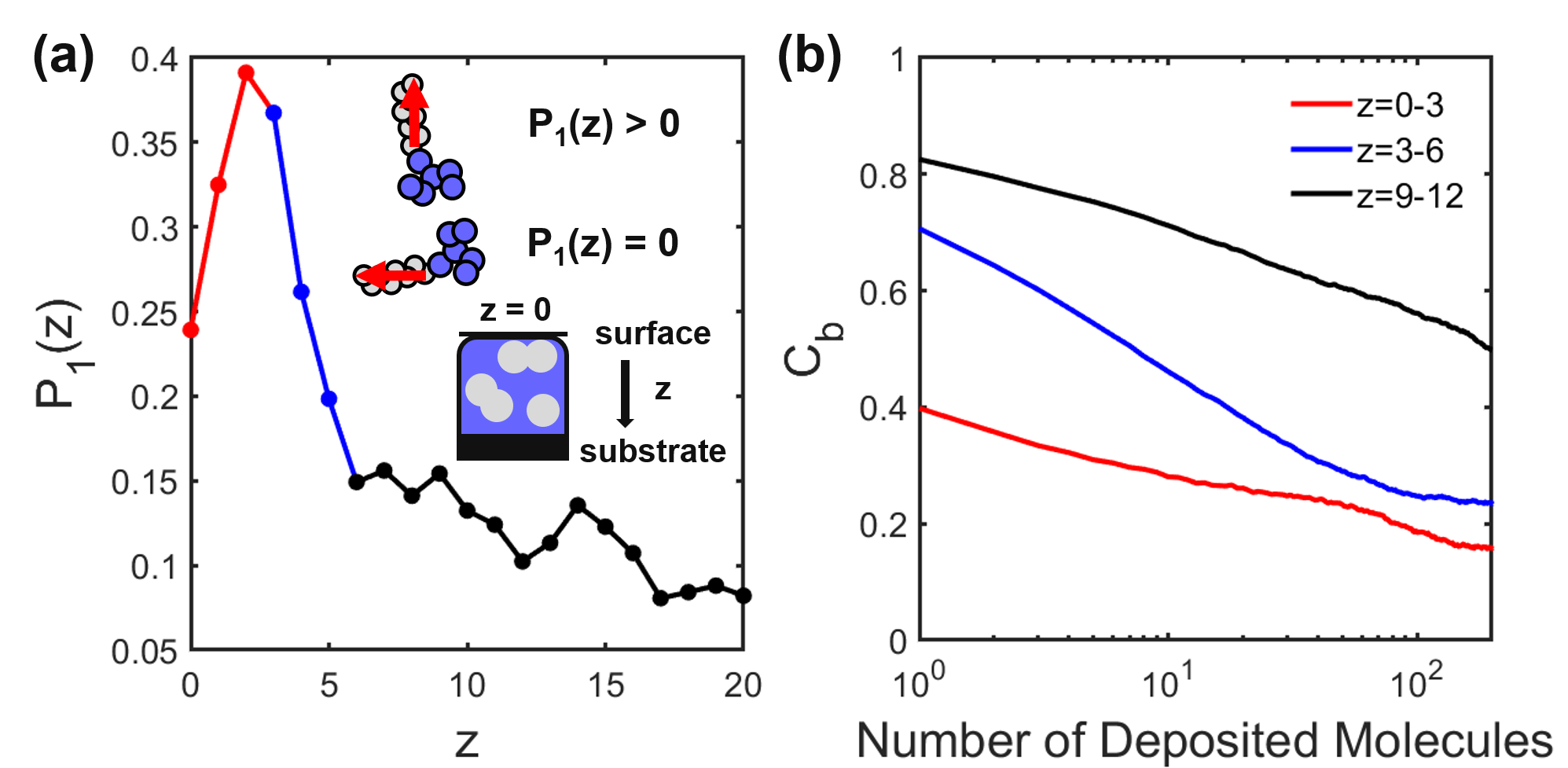}
\caption{(a) $P_1(z)$, orientational order parameter, calculated using the dot product of the end-to-end tail vector with the unit normal to the substrate, as a function of the distance of the center of the tail from the free surface of the film. A positive value indicates that the tail is oriented toward the free surface of the film, and colors represent the surface, secondary, and bulk regions respectively. (b) $P_1$ bond autocorrelation function ($C_b$) of this same tail vector plotted vs. number of subsequent molecules deposited for tails originating in the three distinct regions.}
\label{fig:p1} 
\end{figure}

%%%

The existence of mobility below the free surface in the deposited glasses is further demonstrated by the results in Figure \ref{fig:p1}, which show the orientation of the unit vector along the end-to-end axis of the tail of the 8-tail molecule during the same deposition instance shown in Figure \ref{fig:tail-frac-dep}. Here, to characterize this orientation, we use the $P_1$ orientational order parameter defined by the first Legendre polynomial. $P_1(z)$ is calculated using the dot product of this unit vector, $\textbf{n}(z)$, for a tail a distance $z$ away from the free surface, with the unit normal to the substrate, $\textbf{n}_z$,  
\begin{equation}
P_1(z) = \langle \textbf{n}(z) \cdot \textbf{n}_z \rangle
\label{eq:p1z}
\end{equation}
\noindent such that a positive value of $P_1(z)$ indicates that the tail of the molecule is oriented toward the free surface of the film. Note that the location of the free surface, and thus the origin of this $z$-axis, is evolving throughout the deposition process. Figure \ref{fig:p1}(a) shows that tails in the initial surface layer (red) are the most likely to be oriented toward the free surface during deposition, while those in the secondary layer (blue) are in the process of reorienting to become more isotropic, like those in the bulk of the film (black). Figure \ref{fig:p1}(b) then shows the dynamic behavior of this end-to-end tail vector over the course of a deposition via the $P_1$ bond autocorrelation function, $C_b$, of tails originating at different distances relative to the free surface. Here, the molecular reorientation beyond the initial surface layer can be directly observed. Tails in the surface layer have greater initial mobility, before plateauing after around 50 molecules as they are forced to orient in the positive direction, and then more rapidly reorienting as they enter the secondary layer. Those tails beginning in the secondary layer show a clear reorientation from initial position before then becoming trapped in the bulk-like state. This is a remarkable observation given that previous work has suggested a predominantly surface-diffusion based mechanism for achieving stable PVD glasses. This work clearly demonstrates that relaxation below the free surface plays a role in the ultimate structure of PVD glasses, although it may not necessarily lead to increased stability.

%%%

While differences between surface relaxation dynamics and surface diffusion in small organic molecules have been highlighted in the past \cite{zhang2017decoupling}, few experiments have highlighted the potential role of surface dynamics other than the surface diffusion in the formation of stable glasses \cite{zhang2017invariant}. The clear evidence for rearrangement at layers below the free surface in this study suggest that the surface-mediated equilibration  process may involve more than just the free surface diffusion suggested in the past. Interestingly, this process also hinders layering from the free surface, preventing the formation of bi-layers in the systems chosen in this study, despite significant clustering of the molecules with increased tail length. Thus, we do not observe any longer range structures or domains, and instead we observe disordered clusters.

%%%

\begin{figure}[ht]
\centering
\includegraphics[width=\linewidth]{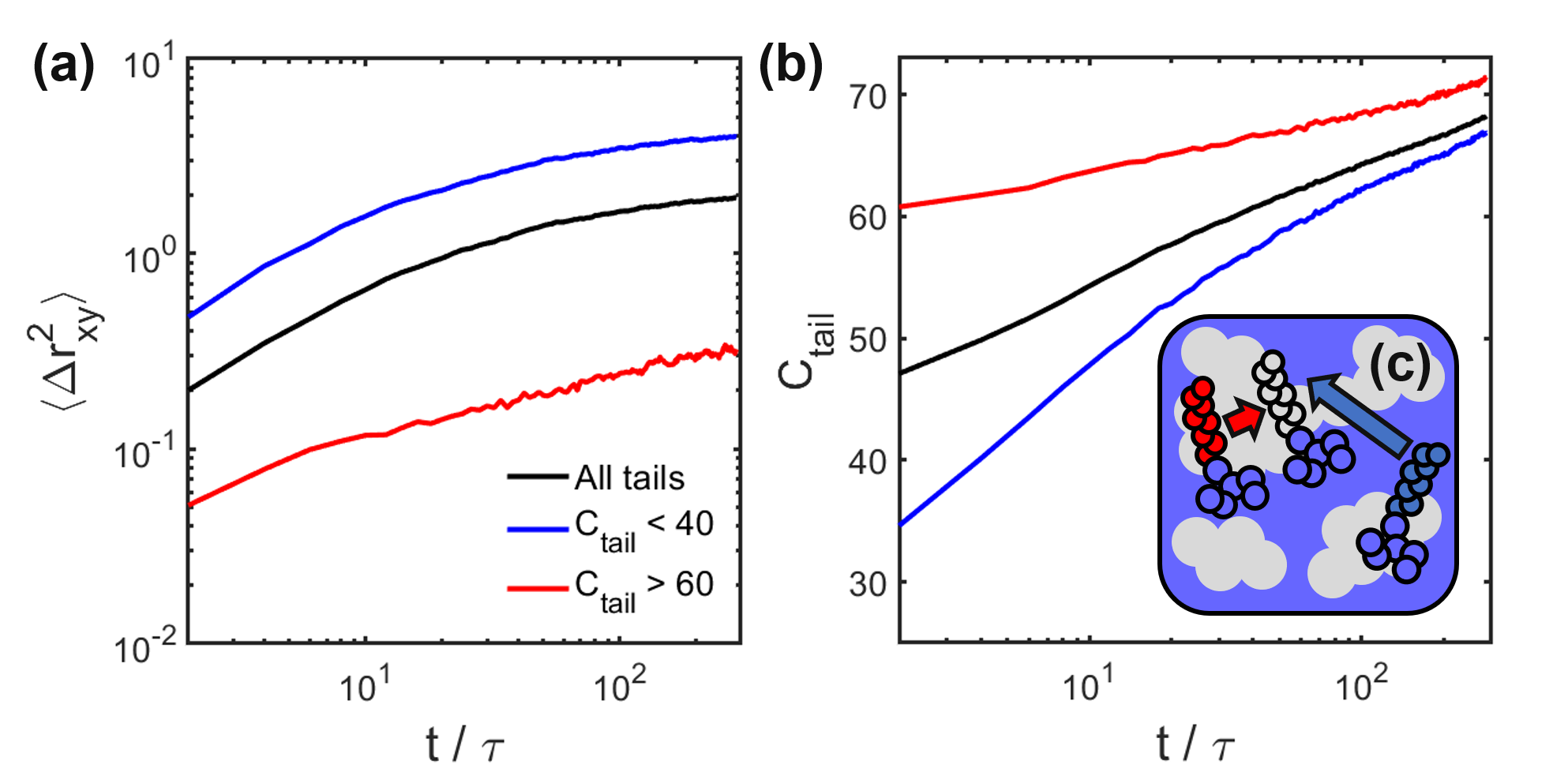}
\caption{(a) MSD in the $x-y$ plane and (b) tail coordination number, $C_\text{tail}$, vs. time, immediately following the initial contact with the film surface for a $300 \tau$ "slow" deposition of 8-tail molecules at $0.90\Tg$. The black line represents all tail particles, the blue line represents those tails for which initially $C_\text{tail} < 40$, and the red line represents those tails for which initially $C_\text{tail} > 60$. (c) A schematic demonstrating the difference in displacement and coordination between tails belonging to these distinct groups. }
\label{fig:coord} 
\end{figure}
 
%%%

\subsection{\label{sec:cluster}The effect of cluster formation on stability of PVD glasses}
Based on the information gleaned from the equilibrium liquid behavior of these molecules, we propose a mechanism for the relative instability of PVD films of molecules with increasing disparity in the interaction potential. The increasingly bulk-like relaxation time and the sub-diffusive surface behavior suggest that as molecules are introduced to the film's free surface, molecules with longer tails will quickly find a configuration in which the "body" and "tail" are matched with like sections, and then become trapped, as they find it increasingly difficult to sample additional configurations. This tendency toward local domain separation interferes with the enhanced mobility typically present at the free surface. Thus, molecules with longer tails will find it more difficult to rearrange into configurations that enable optimal packing, and the resulting PVD glasses will be generally less stable.

%%%

To determine whether this phenomenon held on the free surface of a film during the deposition, we monitored the behavior of the 8-tail molecules in the short time immediately after deposition. To get a sense of how the mobility of these molecules changes with local configuration, we measured the mean squared displacement of the "tail" particles in the $x-y$ plane as well as a tail coordination number, $C_\text{tail}$, defined as the number of tail particles within 1.5$\sbb$ of a tail particle, over the 300 $\tau$ that the molecules were allowed to relax after the initial surface contact. Figure \ref{fig:coord} shows the results of these measurements for all tail particles, and two subgroups: those tails which were initially under-coordinated ($C_\text{tail} < 40$) and those tails which were initially highly coordinated ($C_\text{tail} > 60$). It can be seen that those tails which began in the fairly tail-poor regions exhibit greater than average mobility and evolve toward a greater coordination number much more quickly than other tails. Those which find themselves locally domain separated from the beginning show much lower mobility as well as an overall smaller change in $C_\text{tail}$. Across all "tail" particles in this 8-tail system, the mean squared displacement appears to plateau by the end of the allowed deposition time period. Thus, just as proposed, it seems that these longer tail molecules are either initially mobile and become trapped upon entering a tail-rich region, or, are deposited directly into a locally tail-rich area and are therefore already trapped.

%%%

\begin{figure}[ht]
\centering
\includegraphics[width=\linewidth]{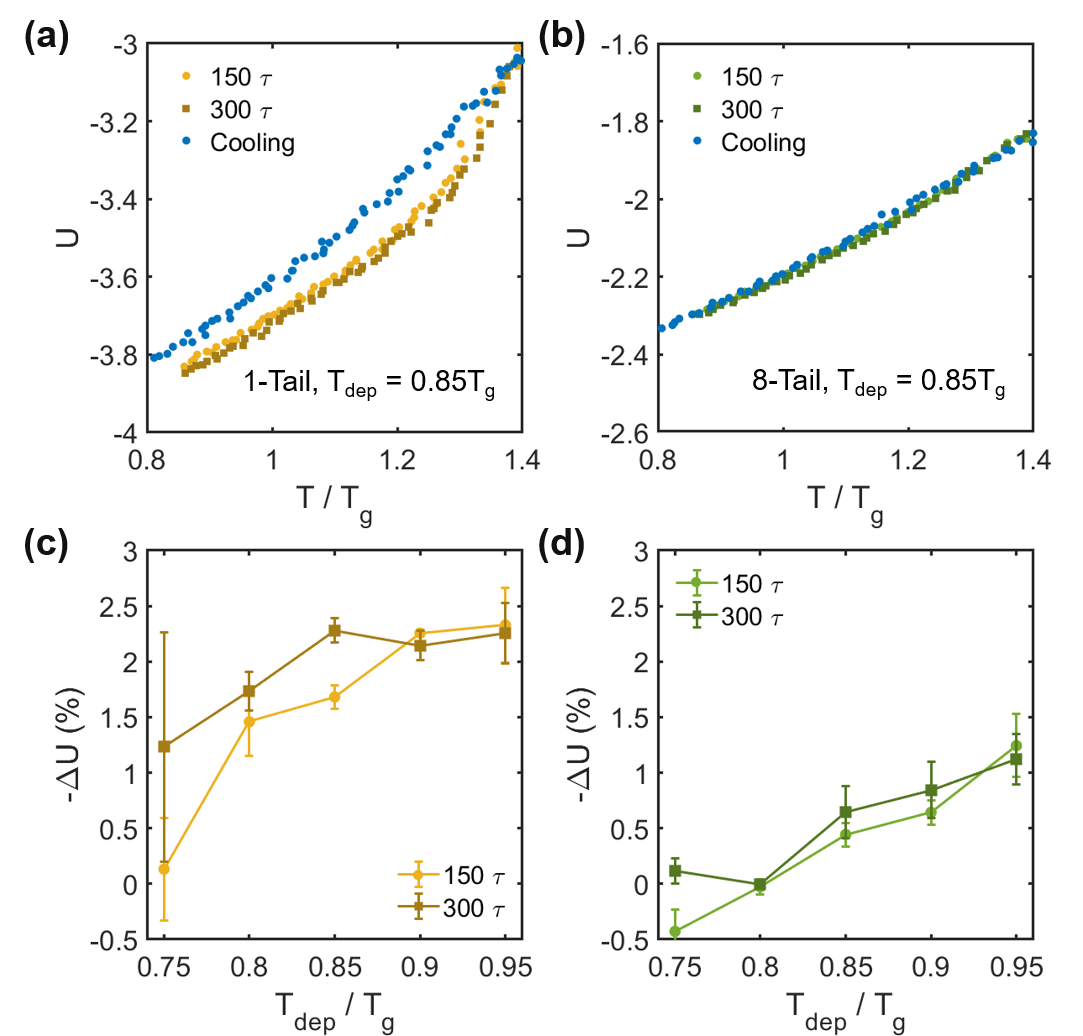}
\caption{(a) Temperature ramp results at $\Tdep = 0.85 \Tg$, plotted as potential energy vs. $T / \Tg$, for a PVD film of 1-tail molecules and (b) 8-tail molecules. (c) Relative potential energy difference between PVD film and ordinary glass across the spectrum of $\Tdep$ for PVD films of 1-tail molecules and (d) 8-tail molecules.}
\label{fig:deprate} 
\end{figure}

%%%

To further verify our proposed mechanism, we then generated additional films of the 1 and 8-tail molecules using the ``slow'' deposition rate across the same range of temperatures, 0.75-0.95$\Tg$. Figure \ref{fig:eqliq} indicates that the 8-tail molecules should not gain much additional mobility from the slower rate, while the 1-tail molecules are generally allowed greater exploration distance. Thus, if the trapping mechanism holds, we would expect that the stability gain for the 1-tail molecule films would be greater than that for the 8-tail molecule films.

%%%

Figure \ref{fig:deprate} demonstrates the observed stability trends for typical vs. slow deposition rates in the form of a temperature ramp simulation and the percent decrease in potential energy across the spectrum of deposition temperatures. Note that we expect the depositions at lower substrate temperatures to exhibit a greater change in potential energy, because at these temperatures kinetics are the limiting factor, as opposed to the higher deposition temperatures, where a thermodynamic limit is quickly imposed by the equilibrium supercooled liquid state. We see that the slower deposition rate made a more significant impact in PVD film stability for the 1-tail molecule films than for 8-tail molecule films. For the 8-tail molecule, all films above the lowest tested substrate temperature show little-to-no difference within our confidence intervals, and for the lowest substrate temperature, the improvement is only from an unstable glass to a film which is nearly identical to one which was liquid cooled. In contrast, the 1-tail molecule films exhibit more dramatic improvement for each of the three lowest tested substrate temperatures. These results are well aligned with the predictions made using the proposed trapping mechanism.

\section{\label{sec:conc}Conclusions}

Here we take a systematic approach to studying the effects of intermolecular interactions and local domain separation on the properties of PVD glasses. This is done using a coarse-grained model of organic glass forming molecules containing fluorocarbon tails of increasing length, which are immiscible with phenyl ``body'' groups. We first show that by increasing the tail length, we increase the tendency for the molecules to form locally separated clusters. We then observe that increasing tail length corresponds to generally less stable PVD glasses, as measured by potential energy change over liquid-cooled glasses and changes in the onset temperature.

%%%

Studying equilibrium liquid films of these molecules provides insight into the behavior of the molecules on the film free surface. Using measurements of MSD in the $x-y$ plane and local relaxation time in the films, we suggest that the longer the tail of the molecule, and thus the larger the disparity in intermolecular interactions, the more likely they are to become locally trapped upon deposition, and therefore unable to optimize packing position. This mechanism is further supported by observations of 8-tail mobility and local configuration at short times immediately after deposition, and simulations at slower deposition rates, which demonstrate that the 8-tail molecules gain little in terms of stability from having a longer time to rearrange. 

%%%

We also observe a strong tendency of the tails to segregate to the free surface as the tail length is increased, resulting in enhanced density of body segments in the layers immediately below the free surface which acts to slow-down surface relaxation times for both equilibrium supercooled liquids and the PVD glasses. However, remarkably, even for the 8-tail molecules the surface relaxation times are fast enough in layers below the free surface that allows molecular reorientation, thus resulting in a constant average tail density throughout the PVD films below this surface layer. As such, despite the strong surface tail segregation, macrostructures such as bi-layers do not form in these systems and only disordered clusters are observed, that can span the size of the system. The enhanced relaxation below the free surface during PVD suggest that stable glass formation involves more than just equilibration due to the free surface diffusion. 

%%%

Understanding the effects of various chemical changes on the surface mobility and packing efficiency of organic glass formers provides us with valuable insight into the fundamentals of this complex process. The mechanism proposed in this study provides a step toward effectively engineering the structure and stability of PVD glass films.

\section{\label{sec:matmet}Materials and Methods}

For the coarse grained models used in this study, each group in the molecule of interest was represented by a sphere interacting with a Lennard-Jones potential truncated and shifted using a linear decay term to ensure the potential and force go continuously to zero at the cutoff. The cutoff distance for the potential is $r_c = 2.5 \sbb$. Substrate particles were fixed in their original positions with a harmonic potential ($k = 50$). Harmonic bonds were placed between each benzene or fluorocarbon pair which are connected in the organic molecule, with lengths adjusted accordingly such that $l = $ 1.0, 0.667 or 0.333 and $k_\text{bond} = 50$. Additionally, appropriate harmonic angles were placed between groups of three particles, with four possible angles ($90^\circ, 120^\circ, 180^\circ, 109.5^\circ, \ k_\text{angle} = 500$).  

%%%

All MD simulations were run using the LAMMPS \cite{plimpton1995fast} package in the NVT ensemble and time step of 0.002. The simulation box was 15$\sbb$ by 15$\sbb$ in the $x-y$ plane and was always allowed at least 10$\sbb$ of vacuum space above the free surface of the film. The PVD process was simulated using a method similar to the one developed by Lyubimov {\it et. al.} utilizing a number of deposition cycles until the film was grown to approximately 50-60$\sbb$ \cite{lyubimov2013model}. A cycle consists of (1) introducing a new, randomly oriented molecule, above the film free surface, (2) linear cooling of the molecule from the high temperature, $T = 1.0$, to $\Tdep$, and (3) minimizing the energy of the system.

%%%

To create the equilibrium liquid, films created via the PVD protocol were heated to well above their glass transition temperature, $T = 1.5T_g$, allowed to relax, and subsequently cooled to $T = 1.2T_g$. Local relaxation time was measured by sorting particles according to their position relative to the $z$-axis, dividing them into bins of equal size, and then fitting the self part of the intermediate scattering function, 
$F_s(\textbf{q},t)$ with $\textbf{q}$ = 7.14, evaluated at $e^{-1}$. 

\begin{acknowledgments}
This work was funded by NSF DMREF grant DMR-1628407 and partially by MRSEC grant,  DMR-1720530. The authors gratefully acknowledge computational resources provided by XSEDE facilities through award TG-DMR150034. The authors thank Georgia Huang for her contribution in designing the molecular structures which are the basis of the coarse grained models used in this work.
\end{acknowledgments}

\bibliography{references}

\end{document}

\section*{Supplementary Information}

\begin{figure}[h!]
\centering
\includegraphics[width=\linewidth]{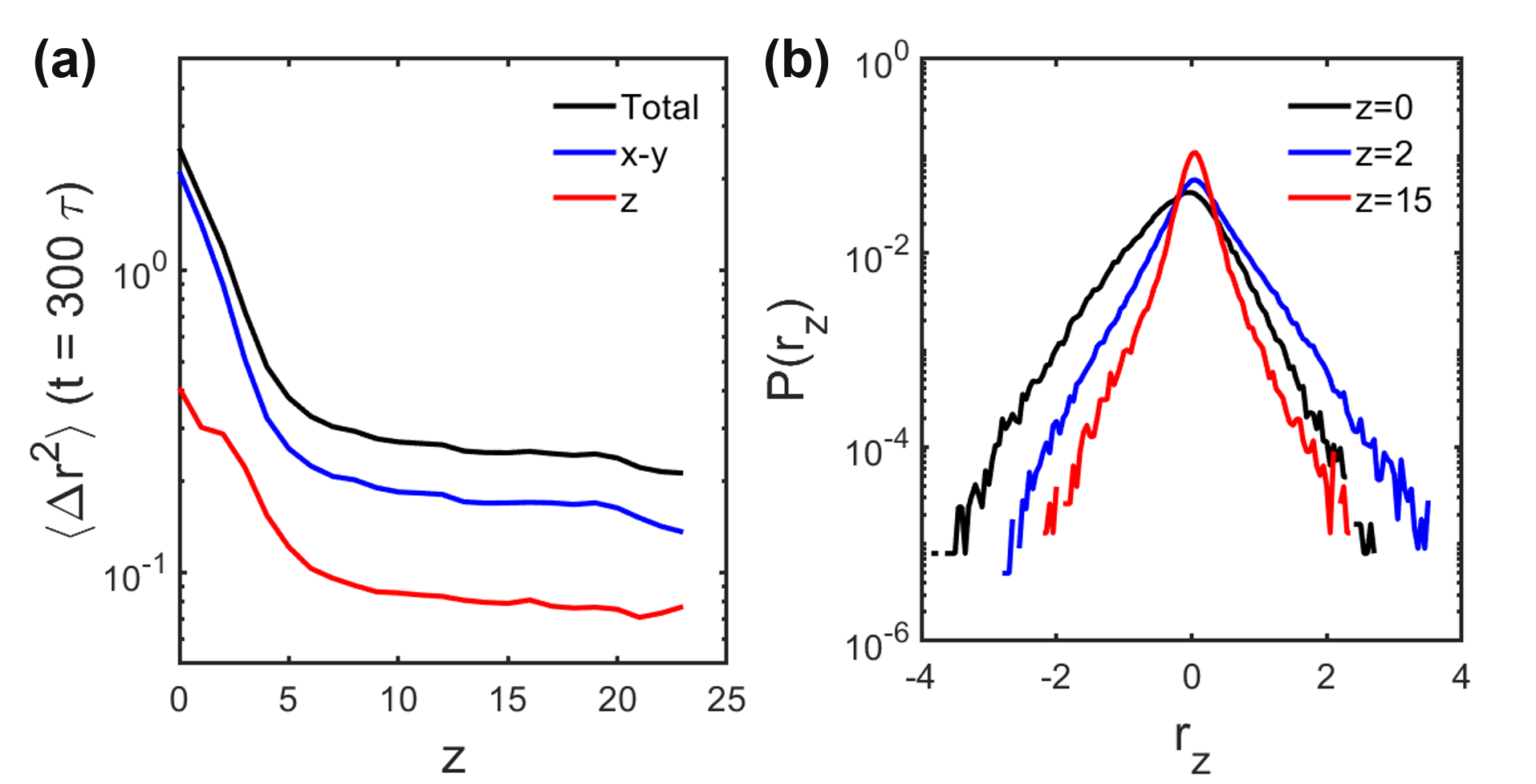}
\caption{(a) Mean squared displacement (MSD) for tail particles during a deposition at $0.90T_g$ as a function of distance away from the free surface, $z$. MSD is given as total, displacement in the $x-y$ plane and displacement along the $z$-axis. (b) Probability distribution of positve (toward the free surface) and negative (toward the substrate) $z$ displacement, $r_z$ for tail particles at the free surface (black), just below the free surface (blue), and in the bulk of the film (red). While the bulk film distribution is practically Gaussian, the free surface distribution is skewed to negative displacements, and the distribution just below the free surface is skewed to positive displacements. Thus, the shoulder in (a) for the $z$ displacement at $z=2$-$3$ represents tail particles actively moving toward the free surface.}
\end{figure}